\documentclass{optica-article}

\journal{opticajournal} 

\articletype{Research Article}

\begin{document}
\title{Refractive Index Tuning of Terahertz Photonic Materials Based on a Stretchable Silicon Effective Medium}
\author{Hidemasa Yamane\authormark{1,*}, Yoshiharu Yamada\authormark{1}, Yusuke Kondo\authormark{1}, Ken Miyajima\authormark{1}, Masayuki Fujita\authormark{2}, and Shuichi Murakami\authormark{1}}

\address{
\authormark{1}Osaka Research Institute of Industrial Science and Technology, 2-7-1 Ayumino, Izumi-city, Osaka, 594-1157, Japan\\
\authormark{2}Graduate School of Engineering Science, The University of Osaka, 1-3 Machikaneyama-cho, Toyonaka, Osaka 560-8531, Japan\\
}

\email{\authormark{*}yamane.hidemasa@orist.jp} 

\begin{abstract*}
Dynamically tunable terahertz (THz) photonics requires low-loss dielectric platforms with practical, continuous control of refractive index. 
Here we present a mechanically reconfigurable THz photonic material platform: a monolithic, all-silicon (Si) stretchable effective medium whose refractive index is tuned by deformation.
A 200 {\textmu}m-thick high-resistivity single-crystal Si slab was patterned into a subwavelength spiral-spring through-hole lattice, rendering bulk Si mechanically compliant while preserving its low-loss dielectric response. 
THz time-domain spectroscopy demonstrates high transmission below $0.6~\mathrm{THz}$ and reveals a monotonic decrease in the effective refractive index under uniaxial stretching. 
At $12.6\%$ elongation, the effective index decreases by 6\% and 8\% for polarizations perpendicular and parallel to the stretch direction, respectively, thereby demonstrating deformation-induced, controllable anisotropy without a detectable increase in extinction. 
This structurally engineered bulk-Si approach offers a process-compatible route to mechanically tunable, low-loss THz components for adaptive wavefront and polarization control.
\end{abstract*}

\section{Introduction}
\label{sec:introduction}

Terahertz (THz) radiation, commonly defined as electromagnetic waves in the 0.1--10~THz range, bridges the gap between microwaves and infrared light and enables non-destructive spectroscopy and imaging.
Because many molecular rotational and vibrational resonances and solid-state excitations lie in this band, THz waves provide contact-free material identification and characterization for applications such as quality inspection, security screening, and biomedical sensing \cite{ferguson2002materials, tonouchi2007cutting}.
In addition, frequencies above 100~GHz (the sub-THz and THz band) are actively considered for 6G-and-beyond wireless links and integrated sensing and communications, where the large available bandwidth can support ultra-high data rates and high-resolution radar functionality \cite{nagatsuma2016advances, akyildiz2014terahertz}.
To fully exploit these opportunities, THz systems require not only efficient sources and detectors but also low-loss, compact, and scalable components for routing, packaging, and wavefront manipulation.

A major challenge at sub-THz and THz frequencies is the development of low-loss components.
Metallic transmission lines and waveguides suffer from increasing ohmic loss due to the skin effect and may exhibit unwanted radiation and crosstalk in dense integration.
These limitations motivate dielectric THz photonics platforms that provide low propagation loss together with wafer-scale manufacturability.
High-resistivity single-crystal silicon (Si) is particularly attractive because it offers low absorption and a high refractive index ($n \approx 3.4$) across a broad THz bandwidth \cite{dai2004terahertz}.
Leveraging mature Si microfabrication, a variety of THz Si photonics components have been developed, including dielectric waveguides and integrated circuits based on photonic-crystal and effective-medium concepts \cite{tsuruda2015extremely, gao2019effective, withayachumnankul2018integrated, koala2022nanophotonics, headland2023terahertz, yamane2025broadband}.
Subwavelength structuring of Si--air composites further enables impedance engineering, tailored effective indices, and birefringence, which are useful for compact lenses and wavefront-control elements.
From a materials perspective, such structuring effectively transforms bulk Si into an engineered THz photonic material with designable effective permittivity and anisotropy.
However, realizing continuous and low-loss tunability in wafer-compatible dielectric platforms remains challenging.

While Si THz photonics has achieved impressive performance in passive components, many emerging systems demand dynamic control of THz phase and wavefront for beam steering, adaptive focusing, and reconfigurable imaging.
A broad range of tunable THz strategies has been explored using carrier-tuned semiconductors, graphene and other two-dimensional materials, phase-change materials, liquid crystals, and MEMS-actuated structures, each with different trade-offs in insertion loss, bandwidth, driving complexity, and stability \cite{zeng2022review}.
Mechanical deformation offers a direct route to tunability, and stretchable THz (meta)surfaces have been demonstrated by patterning resonant elements on elastomeric substrates or embedding dielectric inclusions in stretchable composites \cite{li2013mechanically, lan2018flexible, ambhire2018pattern}.
However, many of these approaches rely on metallic resonances and polymer substrates, which can introduce additional loss and impose limitations in thermal stability, process compatibility, and long-term reliability.

Flexible and stretchable electronics has advanced rapidly to enable devices that conform to curved or soft surfaces.
For Si-based flexible electronics, the dominant approach has been to thin Si into ultrathin membranes and transfer them onto polymer substrates, often combined with serpentine interconnects or mesh layouts to accommodate large deformation \cite{khang2006stretchable, rogers2010materials, kim2008materials}.
Despite their success, polymer substrates can limit dimensional and environmental stability, and thinning and transfer processes may complicate wafer-scale integration.
Conversely, bulk Si wafers are intrinsically brittle and fracture at small tensile deformations, preventing their direct use as flexible photonic substrates.

Here we introduce a mechanically programmable THz photonic material based on structurally flexibilized bulk single-crystal Si.
We form a flexible Si effective medium by MEMS micromachining a subwavelength periodic array of spiral-spring through-holes in a 200 {\textmu}m-thick high-resistivity Si slab.
The spiral-spring unit cell, inspired by mechanical metamaterial concepts \cite{bertoldi2017flexible, rafsanjani2017buckling}, enables large macroscopic deformation while suppressing local stress concentration.
Under uniaxial stretching, the Si--air filling fraction and structural anisotropy evolve continuously, resulting in polarization-dependent tuning of the effective refractive index without introducing additional materials.
Using THz time-domain spectroscopy (THz-TDS), we observe high transmission below 0.6~THz and a deformation-induced effective-index decrease in approximately 6\% (polarization perpendicular to the stretching direction) and up to 8\% (polarization parallel to the stretching direction) at 12.6\% uniaxial elongation.
Full-wave simulations and effective-medium estimates corroborate the measured trends and provide a practical guideline for future reconfigurable, low-loss all-Si THz components.

Although demonstrated here in the THz band, the underlying mechanism is purely geometric and is therefore, in principle, scalable to shorter wavelengths, providing a pathway toward mechanically reconfigurable dielectric meta-optics in the infrared and near-infrared \cite{yu2014flat, khorasaninejad2016metalenses}.

\section{MEMS micromachining for a flexible Si effective medium}
\label{sec:concept}

\subsection{Concept of structural flexibilization of bulk single-crystal Si}
\label{sec:concept_overview}

We realize a flexible and stretchable single-crystal Si substrate by introducing a periodic spring-like network into a bulk Si wafer through MEMS micromachining.
The field of flexible electronics has traditionally relied on the ``make-it-thin'' approach, where Si is thinned down to tens to hundreds of nanometers and then integrated on elastomeric substrates, often combined with serpentine interconnects or island--bridge layouts to accommodate large deformations \cite{khang2006stretchable, rogers2010materials, kim2008materials}.
While effective for enabling mechanical compliance of devices, polymer-based substrates can impose constraints in terms of thermal and dimensional stability and long-term environmental robustness, which become increasingly important when one targets high-precision photonic or high-density electronic integration.

In this work, we pursue a structure-based route: instead of replacing the substrate material, we design bulk single-crystal Si itself as a mechanical metamaterial.
By forming a subwavelength-periodic through-etched pattern, the Si wafer becomes mechanically compliant on the macroscale while remaining a low-loss dielectric for THz waves.
From an electromagnetic viewpoint, the perforated Si/air composite behaves as an effective medium when the lattice period is sufficiently smaller than the wavelength, enabling one to design the effective refractive index through geometry (i.e., the Si/air volume fraction).
From a mechanical viewpoint, a spring-like unit cell allows large global deformation while keeping local strain below the fracture limit of Si.

A key feature of our approach is to directly link the mechanical deformation to the THz effective-medium response.
Specifically, we employ a spiral-spring unit-cell geometry that supports large in-plane uniaxial stretching; upon stretching, the Si/air volume fraction and the structural anisotropy change, resulting in a reversible modulation of the effective refractive index.
This concept is closely related to kirigami-inspired structural metamaterials, where macroscopic mechanical responses are engineered via geometry \cite{rafsanjani2017buckling, bertoldi2017flexible}.
In contrast to mechanically tunable THz metasurfaces that often rely on metals or polymer substrates, our platform aims at all-Si tunability, avoiding additional material loss while maintaining compatibility with established MEMS/Si processes.

\subsection{Design of the Spiral Periodic Spring Structure}

\begin{figure}[t]
  \centering
  \includegraphics[width=\linewidth]{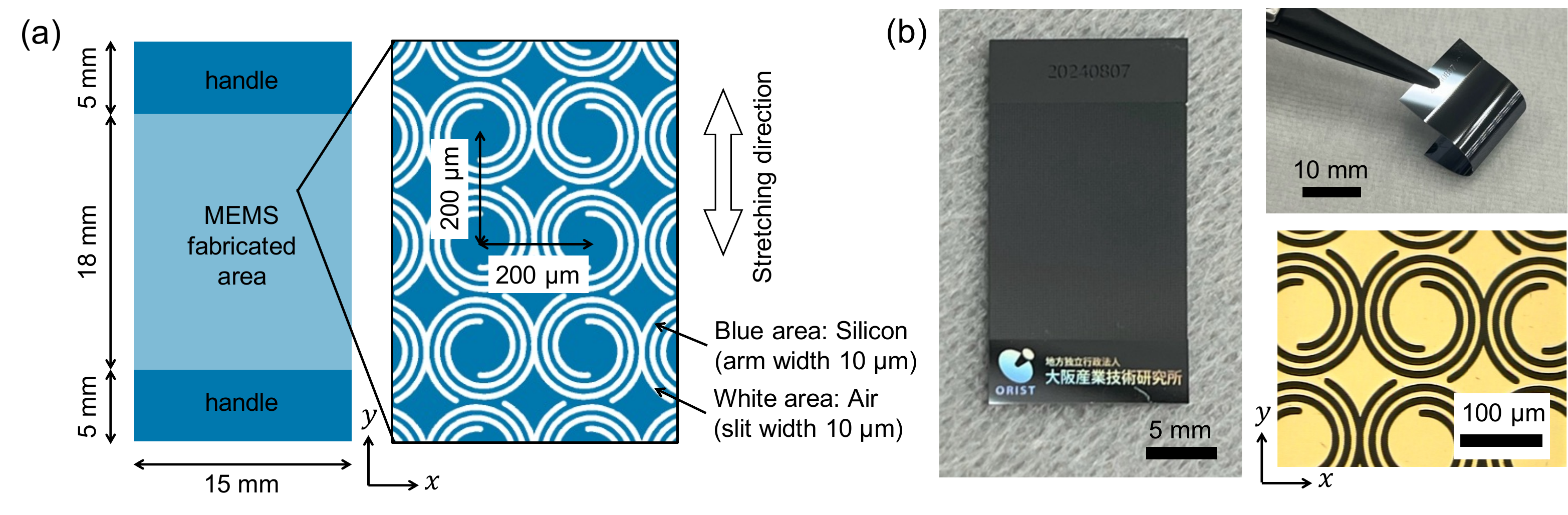}
  \caption{All-Si stretchable effective-medium device based on a spiral-spring through-hole lattice. (a) Schematic of the chip layout (patterned area and rigid handles) and the spiral-spring unit cell with a 200 {\textmu}m pitch and nominal 10 {\textmu}m line/slot width. (b) Photographs of the fabricated 200 {\textmu}m-thick high-resistivity Si chip and an optical micrograph of the spiral-spring lattice.}
  \label{fig:device_photo}
\end{figure}

As shown in Fig.~1(a), the device includes a patterned region of approximately 15 mm $\times$ 18 mm containing the periodic spiral lattice.
The unit cells were arranged on a 200 {\textmu}m pitch in both the $x$ and $y$ directions. 
To realize in-plane uniaxial stretching along the $y$ direction, the unit-cell arrangement was designed to predominantly accommodate deformation along $y$ while suppressing transverse contraction.
The handedness of adjacent spirals (left and right) was alternated to suppress asymmetric deformation under stretching.
As a result, the lattice deforms parallel to the loading direction and can be elongated with an effectively zero Poisson's ratio, i.e., with negligible transverse contraction.
Solid Si handle regions were left at the top and bottom for mechanical gripping. 
A periodic spring structure consisting of spiral through-holes with a line width of 10  {\textmu}m was formed in a 200 {\textmu}m-thick Si plate. 

The periodic spring structure was fabricated using a standard MEMS micromachining process. 
A 200 {\textmu}m-thick high-resistivity single-crystal Si substrate (resistivity $>10~\mathrm{k}\Omega\cdot\mathrm{cm}$) was used. 
A thick photoresist (AZ P4620) was spin-coated and patterned by a maskless direct-write lithography system. 
The pattern was then transferred into the Si substrate by through-wafer deep reactive ion etching (DRIE) using the Bosch process, which alternates SF$_6$ etching and C$_4$F$_8$ passivation steps. 
After etching, the photoresist mask was removed, yielding an all-Si flexible and stretchable structure.

Photographs and optical micrographs of the fabricated device are shown in Fig. \ref{fig:device_photo}(b). 
Optical inspection indicates that the trench width is approximately 11.6 {\textmu}m and the spiral line width is approximately 8.4 {\textmu}m, showing deviations of several micrometers from the nominal 10 {\textmu}m design. These deviations are within typical fabrication tolerances and can be attributed to side-etching during photoresist development and sidewall tapering during the DRIE process. Visualization 1 shows a demonstration of reversible deformation of the fabricated device, highlighting its mechanical stability during handling.

\section{THz time-domain spectroscopy characterization}

\begin{figure}[t]
  \centering
  \includegraphics[width=\linewidth]{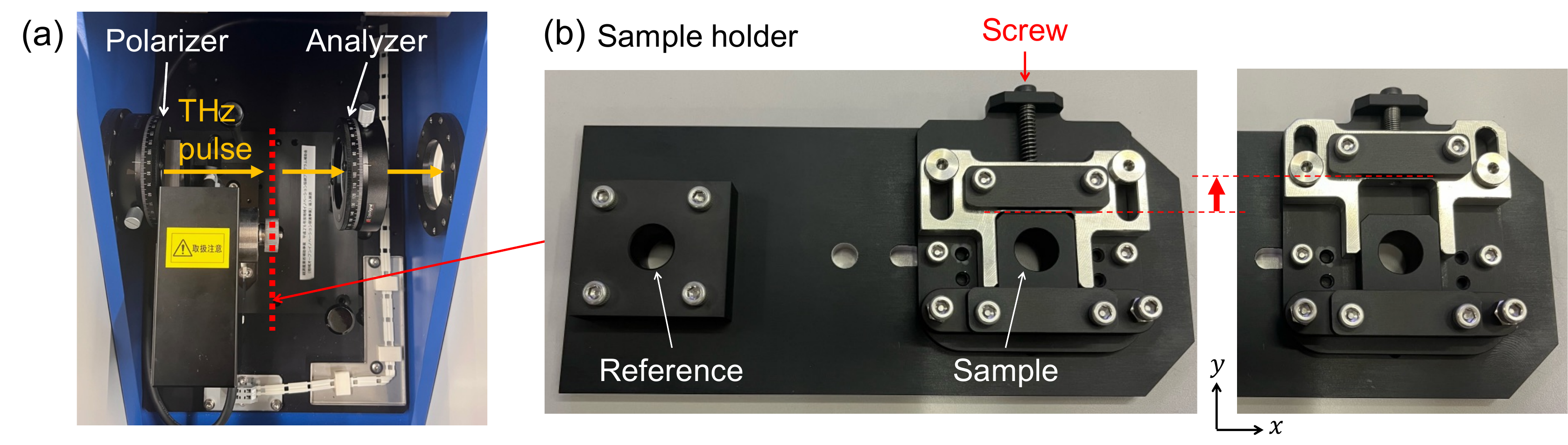}
  \caption{THz-TDS measurement setup and custom stretching fixture.
  (a) Photograph of the transmission configuration in the TeraProspector system (Nippo Precision): the THz pulse is linearly polarized by a polarizer, transmitted through the sample mounted in the sample holder, passed through a second polarizer (analyzer) aligned with the first to measure the co-polarized component, and then detected.
  (b) Photograph of the sample holder (stretching fixture) with two apertures for reference and sample measurements.
  Uniaxial elongation along the in-plane $y$ direction is applied by gripping the bulk-Si handles and tightening the screw.}
  \label{fig:fixture}
\end{figure}

The deformation-dependent THz optical response of the flexible Si effective medium was characterized by THz-TDS in a standard transmission configuration using a commercial system (TeraProspector, Nippo Precision) \cite{jepsen2011terahertz, hangyo2002spectroscopy}.
As shown in Fig.~\ref{fig:fixture}(a), the generated THz pulse was linearly polarized by a polarizer, transmitted through the sample mounted in the sample holder, and then passed through a second polarizer (analyzer) before detection.
A photograph of the sample holder is shown in Fig.~\ref{fig:fixture}(b).
The sample was mounted in an aperture with a diameter of 10~mm such that the patterned effective-medium region fully covered the THz beam.
The collimated THz beam was spatially filtered by the aperture and formed an approximately Gaussian spot at the sample position.

After a 30-min purge with dry air, for each elongation condition, we alternated between sample and reference acquisitions, recording 40 time-domain waveforms for each, and averaging them to improve the signal-to-noise ratio.
The spectra of the transmission intensity and the phase delay were obtained by applying a Fourier transform to the measured time-domain waveform.
The frequency resolution was approximately 5.7~GHz.

In our setup, the THz beam diameter was on the order of three times the free-space wavelength.
In the frequency range of interest ($\le 0.6$~THz), the beam diameter was typically $\gtrsim 1.5$~mm, which is much larger than the unit-cell pitch (200 {\textmu}m).
Therefore, the measured transmission represents an average response over many unit cells and can be treated as an effective-medium response.

Mechanical stretching was applied using the sample holder by gripping the bulk-Si handle regions.
As illustrated in Fig.~\ref{fig:fixture}(b), tightening the screw at the clamping part pulls the chip along the in-plane $y$ direction, enabling controlled uniaxial elongation.
The uniaxial stretching procedure using the custom fixture shown in Fig. 2(b) is demonstrated in Visualization 2.

For each elongation condition, the averaged complex transmission was analyzed using a thin-slab transmission model \cite{hangyo2002spectroscopy} to extract the effective refractive index.
Specifically, the patterned Si effective medium was modeled as a homogeneous freestanding thin film with thickness $d=$ 200 {\textmu}m, and multiple internal reflections at the air/film interfaces were included (Fabry--P\'erot effect).
The effective refractive index was determined by fitting the model transmission to the measured complex spectrum.

\section{Results}

\subsection{Stretch-induced refractive-index control}
\label{sec:trans_phase}

\begin{figure}[t]
  \centering
  \includegraphics[width=0.8\linewidth]{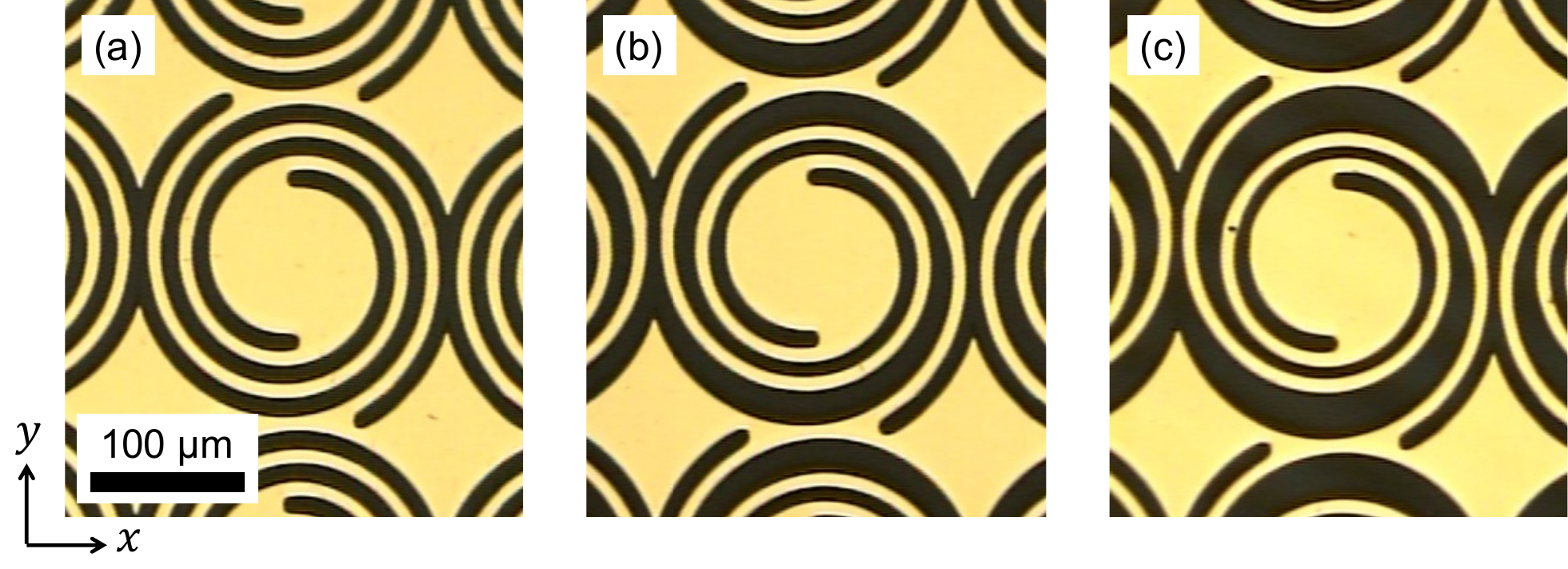}
  \caption{Optical micrographs of the spiral-spring effective-medium region under (a) relaxed (natural-length), (b) 7.3\% elongation, and (c) 12.6\% elongation along the $y$ direction.
  The dark regions correspond to through-etched air gaps, which broaden with increasing elongation.}
  \label{fig:micropic}
\end{figure}

Optical micrographs of the patterned region under different elongation states are shown in Fig.~\ref{fig:micropic}:
(a) the relaxed (natural-length) state, (b) and (c) nominal uniaxial elongations of 7.3\% and 12.6\% along the $y$ axis, respectively.
The dark regions correspond to the through-etched gaps (air regions), which widen as the elongation increases.

\begin{figure}[tphb]
  \centering
  \includegraphics[width=\linewidth]{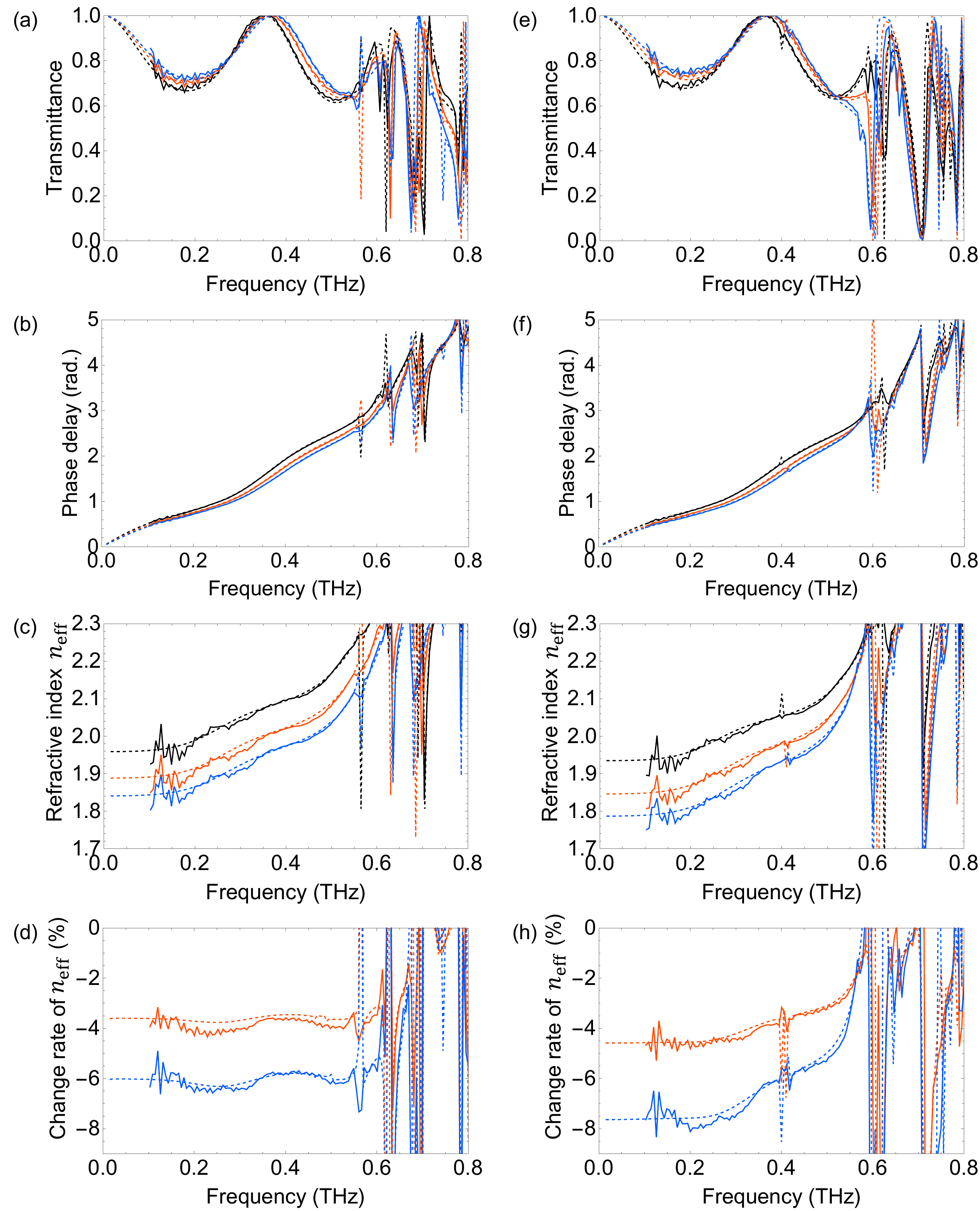}
  \caption{Measured (solid) and simulated (dashed) spectra and retrieved effective refractive index for different elongation ratios.
  (a)--(d) $x$-polarized incidence (perpendicular to the stretching direction): (a) transmittance, (b) phase delay, (c) retrieved effective refractive index, and (d) relative change in refractive index referenced to the relaxed state.
  (e)--(h) $y$-polarized incidence (parallel to the stretching direction): (e) transmittance, (f) phase delay, (g) retrieved effective refractive index, and (h) relative change.
  Black: relaxed; red: 7.3\% elongation; blue: 12.6\% elongation.}
  \label{fig:result}
\end{figure}

Upon stretching, the originally quasi-isotropic spiral-spring unit cell deforms anisotropically, resulting in a reduced Si filling fraction in the effective-medium region.
Because the structure consists of through-wafer holes, the areal fraction estimated from the planar optical micrographs can be used as an approximation for the volumetric filling fraction.
From image-based estimation of the Si and air areas in Fig.~\ref{fig:micropic}, we obtained the Si filling fraction $f_{\mathrm{Si}}$ (and the air filling fraction $f_{\mathrm{air}}=1-f_{\mathrm{Si}}$).
In the relaxed state, $f_{\mathrm{Si}}\approx0.60$.
With 7.3\% elongation, $f_{\mathrm{Si}}\approx0.56$, and with 12.6\% elongation, $f_{\mathrm{Si}}\approx0.53$.
This increase in the air fraction reduces the effective permittivity of the Si--air composite and is a primary origin of the decrease in the effective refractive index $n_{\mathrm{eff}}$.

THz transmission measurements were performed for the relaxed state and for uniaxial elongations of 7.3\% and 12.6\% along the $y$ axis.
Because stretching induces optical anisotropy, we measured the transmission for incident polarizations along both $x$ and $y$, where $x$ is perpendicular to the stretching direction and $y$ is parallel to it.
The measured and simulated results are summarized in Fig.~\ref{fig:result}.
The black curves correspond to the relaxed state, and the red and blue curves correspond to 7.3\% and 12.6\% elongation, respectively.
Solid curves denote experiments, and dashed curves denote full-wave electromagnetic simulations using COMSOL Multiphysics.
Because the 10-mm aperture in the sample holder limits the measurable low-frequency response to approximately 0.1~THz, the experimental data are plotted for frequencies above 0.1~THz.

Figures~\ref{fig:result}(a), (b) show the transmittance and phase-delay spectra for $x$-polarized incidence, obtained from the Fourier transforms of the measured time-domain waveforms.
High transmittance characteristic of high-resistivity Si is maintained.
In the frequency range below 0.6~THz, Fabry--P\'erot-type interference fringes consistent with a homogeneous effective slab are observed, supporting an effective-medium interpretation.
Above 0.6~THz, the simple thin-slab interference pattern is disrupted; nevertheless, the measured spectra remain in good agreement with the simulations, and additional spectral features attributable to the periodic microstructure are observed.
From these spectra, the effective refractive-index spectra were extracted, as shown in Fig.~\ref{fig:result}(c).
Because the device consists only of high-resistivity Si and air, the extracted extinction coefficient was found to be very small within the experimental sensitivity, and no systematic increase in loss was observed upon stretching.

The effective refractive index decreases monotonically with elongation.
Figure~\ref{fig:result}(d) plots the relative change in refractive index referenced to the relaxed state.
For $x$ polarization (perpendicular to stretching), the relative change exhibits little frequency dependence, and $n_{\mathrm{eff}}$ decreases by approximately 6\% at 12.6\% elongation.

The corresponding results for $y$-polarized incidence (parallel to stretching direction) are shown in Figs.~\ref{fig:result}(e)--(h).
As in the $x$-polarized case, high transmittance is maintained while the effective refractive index decreases with elongation.
The relative reduction reaches approximately 8\%; however, unlike the $x$-polarized case, the refractive-index modulation shows a clearer frequency dependence.
This behavior is attributed to deformation-induced anisotropy, including the change in the structural period along the stretching direction.

\subsection{Consistency with effective-medium estimates}
\label{sec:emt_consistency}
In the previous subsection, we showed that the effective refractive index $n_{\mathrm{eff}}$ extracted from the THz-TDS measurements is well reproduced by full-wave electromagnetic simulations based on the finite-element method.
In this subsection, we focus on the frequency range below $0.6~\mathrm{THz}$, where the effective-medium interpretation is valid, and demonstrate order-of-magnitude consistency with the Maxwell--Garnett (MG) approximation.
This comparison further supports that the retrieved $n_{\mathrm{eff}}$ can be consistently understood as the homogenized response of a Si--air composite.
According to the MG approximation \cite{garnett1904xii, zhang2015effective, gao2019effective},
\begin{equation}
\varepsilon_{\mathrm{eff}} =
\varepsilon_{\mathrm{air}}
\frac{\varepsilon_{\mathrm{Si}} + 2\varepsilon_{\mathrm{air}} + 2f_{\mathrm{Si}}\left(\varepsilon_{\mathrm{Si}}-\varepsilon_{\mathrm{air}}\right)}
{\varepsilon_{\mathrm{Si}} + 2\varepsilon_{\mathrm{air}} - f_{\mathrm{Si}}\left(\varepsilon_{\mathrm{Si}}-\varepsilon_{\mathrm{air}}\right)} ,
\label{eq:MG}
\end{equation}
where $f_{\mathrm{Si}}$ is the Si filling fraction, $\varepsilon_{\mathrm{Si}} = n_{\mathrm{Si}}^2$, and $\varepsilon_{\mathrm{air}} \approx 1$.
High-resistivity single-crystal silicon exhibits an almost frequency-independent refractive index of $n_{\mathrm{Si}} \approx 3.4$ in the relevant THz band \cite{dai2004terahertz}.
In the present MG estimate, in-plane anisotropy (i.e., polarization-dependent differences) is not considered, and the comparison is limited to demonstrating the order of magnitude of the average effective permittivity.

The decrease in $n_{\mathrm{eff}}$ with stretching can be understood from an effective-medium perspective.
Specifically, stretching increases the air fraction in the silicon--air composite, thereby reducing the effective permittivity.
From optical micrographs, the silicon filling fraction was estimated to be $f_{\mathrm{Si}} \approx 0.60$ in the relaxed state and to decrease to $f_{\mathrm{Si}} \approx 0.53$ at 12.6\% elongation.
Substituting $n_{\mathrm{Si}} \approx 3.4$ into Eq.~(\ref{eq:MG}), the corresponding effective refractive index $n_{\mathrm{eff}}(= \sqrt{\varepsilon_{\mathrm{eff}}})$ is estimated to decrease from 1.91 to 1.76, i.e., by approximately 8\%.
This estimate is consistent, in order of magnitude, with the measured refractive-index reduction of 6--8\%.
These results support the validity of the effective-medium approximation in the low-frequency regime, where the spiral-spring periodicity is sufficiently subwavelength.


\subsection{Structural Mechanical Analysis of the Spiral Periodic Spring Structure}
\begin{figure}[t]
  \centering
  \includegraphics[width=0.45\linewidth]{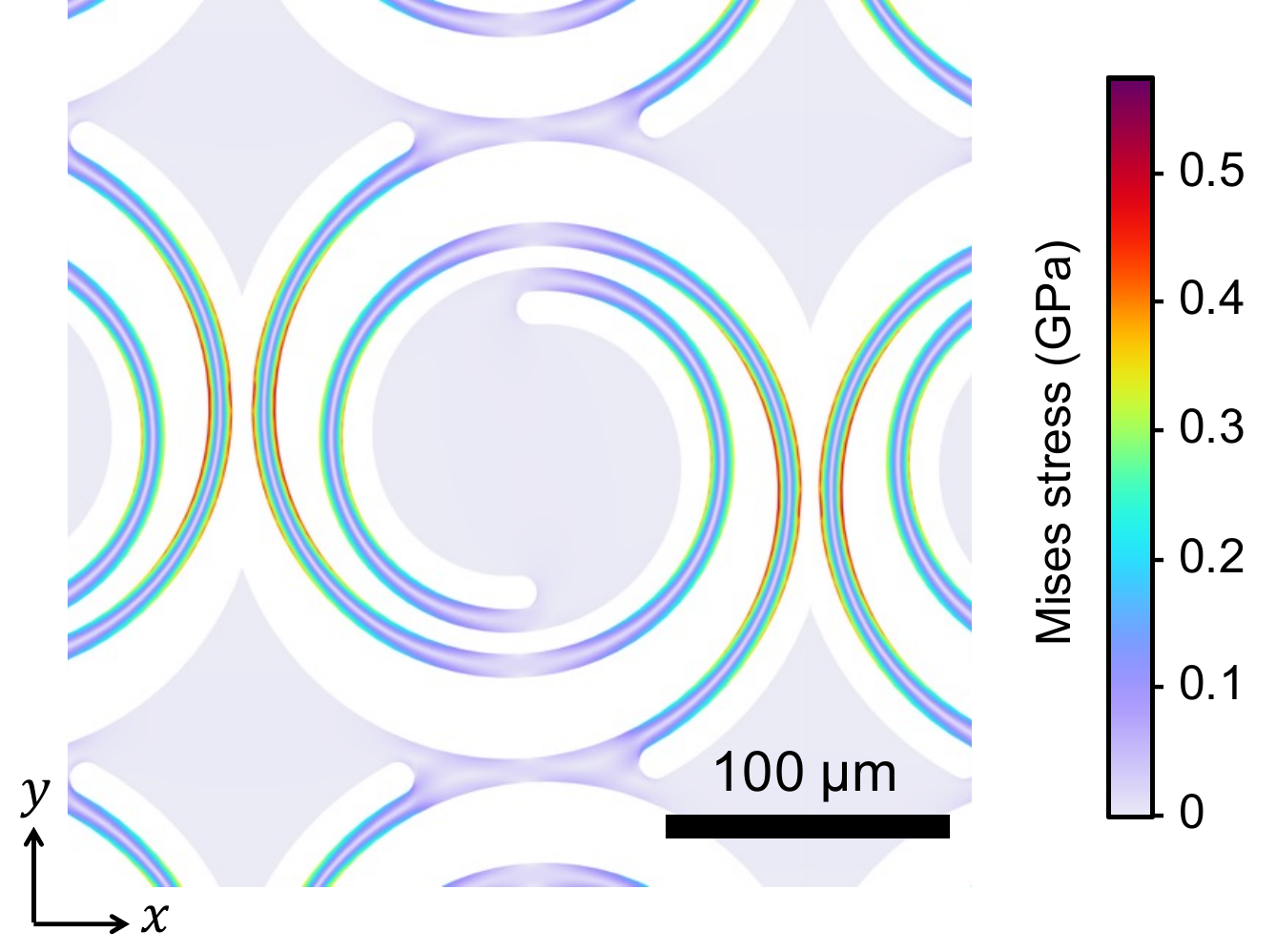}
  \caption{FEM-simulated von Mises stress distribution in the spiral-spring periodic structure under 12.6\% uniaxial elongation along the $y$ direction. The stress hotspot appears around the midspan of the outermost spiral arm, reaching a peak value of 0.57~GPa.}
  \label{fig:stretching_fixture}
\end{figure}

To evaluate mechanical reliability and to guide geometry optimization, finite element method (FEM) simulations were performed using COMSOL Multiphysics on a representative volume element consisting of multiple unit cells. The spiral through-etched spring pattern used in the simulation is shown in Fig. \ref{fig:stretching_fixture}.
Si was modeled as an isotropic linear elastic material with a Young's modulus of $E=170~\mathrm{GPa}$ and a Poisson's ratio of $\nu=0.28$. When a global elongation of approximately $12.6\%$ was applied, the maximum stress was found to concentrate near the outer perimeter of the spiral springs, reaching approximately $0.57~\mathrm{GPa}$.

As seen in Fig.~\ref{fig:stretching_fixture}, the highest stress is localized around the midspan of the outermost spiral arm, indicating that local bending of this arm dominates the stress concentration under global elongation.


The fracture strength of micromachined single-crystal Si exhibits significant variability depending on surface defects, specimen dimensions, crystallographic orientation, and etching-induced damage.
In the present spiral-spring geometry, the curved arms span a range of in-plane directions; therefore, the local tensile axis is not restricted to a single crystallographic direction.
Accordingly, we compare the simulated peak stress with representative fracture-strength ranges reported for micromachined single-crystal Si microstructures.
Typical measured strengths for micro- and nano-scale Si components are on the order of 1--5~$\mathrm{GPa}$, while DRIE-fabricated specimens commonly show gigapascal-level strength distributions depending on etch-induced surface morphology \cite{delrio2015fracture,gaither2013etching}.

Compared with these representative ranges, the simulated peak von Mises stress of $0.57~\mathrm{GPa}$ is lower than typical reported fracture strengths, suggesting a practical safety margin for elongations on the order of 12.6\% in the present design.
A more rigorous reliability assessment would require strength statistics specific to the present fabrication process and geometry \cite{gaither2013etching}.

Beyond reliability, the stress map in Fig.~\ref{fig:stretching_fixture} provides a concrete design target for stress redistribution. A key next step is to combine FEM-based mechanical screening with full-wave electromagnetic-response analysis to redesign the spring topology so as to reduce the peak stress at the hotspot and realize a more compliant unit cell that accommodates larger macroscopic elongation. Such multiphysics co-design is expected to expand the mechanically programmable THz response, enabling a larger deformation-induced change in the effective refractive index together with stronger, tunable anisotropy.

\section{Summary and Conclusion}
\label{sec:conclusion}

We demonstrated a monolithic, all-Si stretchable effective medium for THz waves by MEMS micromachining a 200 {\textmu}m-thick high-resistivity single-crystal Si slab into a subwavelength array of spiral-spring through-holes.
The spiral-spring geometry converts brittle bulk Si into a mechanically compliant ``mechanical metamaterial,'' while uniaxial stretching modifies the Si/air filling fraction and induces structural anisotropy, enabling refractive-index tuning without material substitution.

THz-TDS measurements revealed high transmission below 0.6~THz with Fabry--P\'erot fringes consistent with a homogeneous effective-slab response.
By fitting the measured complex transmission with a multiple-reflection thin-slab model, we retrieved a deformation-dependent effective refractive index that decreases monotonically with elongation.
At 12.6\% elongation, $n_{\mathrm{eff}}$ decreases by 6\% for polarization perpendicular to the stretching direction ($x$) and by up to 8\% for polarization parallel to it ($y$).
The distinct tuning for the two orthogonal polarizations indicates mechanically induced and controllable birefringence, suggesting a straightforward extension to polarization-control functionalities.
Effective-medium estimates based on micrograph-derived filling fractions ($f_{\mathrm{Si}}: 0.60 \rightarrow 0.53$) predict an index reduction of 8\%, consistent with the measured 6--8\% modulation and full-wave simulations.
Within the experimental sensitivity, no systematic increase in extinction was observed upon stretching.
FEM analysis indicates a peak von Mises stress of 0.57~GPa at 12.6\% deformation, which is lower than typical fracture strengths reported for micromachined single-crystal Si (on the order of 1--5~$\mathrm{GPa}$), suggesting a practical safety margin for the demonstrated elongation \cite{delrio2015fracture,gaither2013etching}.

These results establish structurally flexibilized bulk Si as a process-compatible route to mechanically reconfigurable, low-loss THz effective media.
Future work will combine FEM-based mechanical optimization and full-wave electromagnetic co-design to explore spring geometries that redistribute stress and accommodate larger elongation while enhancing refractive-index tunability and deformation-induced birefringence.
Such mechanically programmable anisotropic effective media would enable not only index tuning but also tunable polarization optics and spatially programmed wavefront-shaping components, including reconfigurable dielectric metasurfaces and metalenses in the THz regime \cite{kildishev2013planar, yu2014flat, khorasaninejad2016metalenses, jiang2018all}.
Importantly, because the concept relies on purely geometric subwavelength structuring of silicon, it is in principle scalable to shorter wavelengths, potentially extending beyond the THz band to the infrared and near-infrared by shrinking the lattice pitch and thickness.
Realizing such mechanically reconfigurable dielectric meta-optics in the NIR would require submicron and nanoscale fabrication and a corresponding redesign of the spring topology (e.g., using thin Si layers such as SOI membranes) to maintain mechanical compliance.


\begin{backmatter}
\bmsection{Funding}
JSPS KAKENHI (Grant Number: JP24K00933).

\bmsection{Acknowledgment}
This work was supported in part by JSPS KAKENHI (Grant Number: JP24K00933).

\bmsection{Disclosures}
The authors declare no conflicts of interest.

\bmsection{Data Availability Statement}
Data underlying the results presented in this paper are not publicly available at this time but may be obtained from the authors upon reasonable request.

\end{backmatter}

\bibliography{sample}

\begin{thebibliography}{10}
\newcommand{\enquote}[1]{``#1''}

\bibitem{ferguson2002materials}
B.~Ferguson and X.-C. Zhang, \enquote{Materials for terahertz science and
  technology,} {\protect\JournalTitle{Nature materials}} \textbf{1}, 26--33
  (2002).

\bibitem{tonouchi2007cutting}
M.~Tonouchi, \enquote{Cutting-edge terahertz technology,}
  {\protect\JournalTitle{Nature photonics}} \textbf{1}, 97--105 (2007).

\bibitem{nagatsuma2016advances}
T.~Nagatsuma, G.~Ducournau, and C.~C. Renaud, \enquote{Advances in terahertz
  communications accelerated by photonics,} {\protect\JournalTitle{Nature
  Photonics}} \textbf{10}, 371--379 (2016).

\bibitem{akyildiz2014terahertz}
I.~F. Akyildiz, J.~M. Jornet, and C.~Han, \enquote{Terahertz band: Next
  frontier for wireless communications,} {\protect\JournalTitle{Physical
  communication}} \textbf{12}, 16--32 (2014).

\bibitem{dai2004terahertz}
J.~Dai, J.~Zhang, W.~Zhang, and D.~Grischkowsky, \enquote{Terahertz time-domain
  spectroscopy characterization of the far-infrared absorption and index of
  refraction of high-resistivity, float-zone silicon,}
  {\protect\JournalTitle{Journal of the Optical Society of America B}}
  \textbf{21}, 1379--1386 (2004).

\bibitem{tsuruda2015extremely}
K.~Tsuruda, M.~Fujita, and T.~Nagatsuma, \enquote{Extremely low-loss terahertz
  waveguide based on silicon photonic-crystal slab,}
  {\protect\JournalTitle{Optics express}} \textbf{23}, 31977--31990 (2015).

\bibitem{gao2019effective}
W.~Gao, X.~Yu, M.~Fujita, \emph{et~al.}, \enquote{Effective-medium-cladded
  dielectric waveguides for terahertz waves,} {\protect\JournalTitle{Optics
  express}} \textbf{27}, 38721--38734 (2019).

\bibitem{withayachumnankul2018integrated}
W.~Withayachumnankul, M.~Fujita, and T.~Nagatsuma, \enquote{Integrated silicon
  photonic crystals toward terahertz communications,}
  {\protect\JournalTitle{Advanced Optical Materials}} \textbf{6}, 1800401
  (2018).

\bibitem{koala2022nanophotonics}
R.~A. Koala, M.~Fujita, and T.~Nagatsuma, \enquote{Nanophotonics-inspired
  all-silicon waveguide platforms for terahertz integrated systems,}
  {\protect\JournalTitle{Nanophotonics}} \textbf{11}, 1741--1759 (2022).

\bibitem{headland2023terahertz}
D.~Headland, M.~Fujita, G.~Carpintero, \emph{et~al.}, \enquote{Terahertz
  integration platforms using substrateless all-silicon microstructures,}
  {\protect\JournalTitle{APL Photonics}} \textbf{8} (2023).

\bibitem{yamane2025broadband}
H.~Yamane, Y.~Yamada, Y.~Kondo, \emph{et~al.}, \enquote{Broadband terahertz
  half-wave plate based on an all-silicon anisotropic effective medium,}
  {\protect\JournalTitle{Optics Express}} \textbf{33}, 53279--53294 (2025).

\bibitem{zeng2022review}
H.~Zeng, S.~Gong, L.~Wang, \emph{et~al.}, \enquote{A review of terahertz phase
  modulation from free space to guided wave integrated devices,}
  {\protect\JournalTitle{Nanophotonics}} \textbf{11}, 415--437 (2022).

\bibitem{li2013mechanically}
J.~Li, C.~M. Shah, W.~Withayachumnankul, \emph{et~al.}, \enquote{Mechanically
  tunable terahertz metamaterials,} {\protect\JournalTitle{Applied Physics
  Letters}} \textbf{102} (2013).

\bibitem{lan2018flexible}
C.~Lan, D.~Zhu, J.~Gao, \emph{et~al.}, \enquote{Flexible and tunable terahertz
  all-dielectric metasurface composed of ceramic spheres embedded in
  ferroelectric/elastomer composite,} {\protect\JournalTitle{Optics express}}
  \textbf{26}, 11633--11638 (2018).

\bibitem{ambhire2018pattern}
S.~Ambhire, S.~Palkhivala, A.~Agrawal, \emph{et~al.}, \enquote{“pattern and
  peel” method for fabricating mechanically tunable terahertz metasurface on
  an elastomeric substrate,} {\protect\JournalTitle{Optical Materials Express}}
  \textbf{8}, 3382--3391 (2018).

\bibitem{khang2006stretchable}
D.-Y. Khang, H.~Jiang, Y.~Huang, and J.~A. Rogers, \enquote{A stretchable form
  of single-crystal silicon for high-performance electronics on rubber
  substrates,} {\protect\JournalTitle{Science}} \textbf{311}, 208--212 (2006).

\bibitem{rogers2010materials}
J.~A. Rogers, T.~Someya, and Y.~Huang, \enquote{Materials and mechanics for
  stretchable electronics,} {\protect\JournalTitle{science}} \textbf{327},
  1603--1607 (2010).

\bibitem{kim2008materials}
D.-H. Kim, J.~Song, W.~M. Choi, \emph{et~al.}, \enquote{Materials and
  noncoplanar mesh designs for integrated circuits with linear elastic
  responses to extreme mechanical deformations,}
  {\protect\JournalTitle{Proceedings of the National Academy of Sciences}}
  \textbf{105}, 18675--18680 (2008).

\bibitem{bertoldi2017flexible}
K.~Bertoldi, V.~Vitelli, J.~Christensen, and M.~Van~Hecke, \enquote{Flexible
  mechanical metamaterials,} {\protect\JournalTitle{Nature Reviews Materials}}
  \textbf{2}, 1--11 (2017).

\bibitem{rafsanjani2017buckling}
A.~Rafsanjani and K.~Bertoldi, \enquote{Buckling-induced kirigami,}
  {\protect\JournalTitle{Physical review letters}} \textbf{118}, 084301 (2017).

\bibitem{yu2014flat}
N.~Yu and F.~Capasso, \enquote{Flat optics with designer metasurfaces,}
  {\protect\JournalTitle{Nature materials}} \textbf{13}, 139--150 (2014).

\bibitem{khorasaninejad2016metalenses}
M.~Khorasaninejad, W.~T. Chen, R.~C. Devlin, \emph{et~al.}, \enquote{Metalenses
  at visible wavelengths: Diffraction-limited focusing and subwavelength
  resolution imaging,} {\protect\JournalTitle{Science}} \textbf{352},
  1190--1194 (2016).

\bibitem{jepsen2011terahertz}
P.~U. Jepsen, D.~G. Cooke, and M.~Koch, \enquote{Terahertz spectroscopy and
  imaging--modern techniques and applications,} {\protect\JournalTitle{Laser \&
  Photonics Reviews}} \textbf{5}, 124--166 (2011).

\bibitem{hangyo2002spectroscopy}
M.~Hangyo, T.~Nagashima, and S.~Nashima, \enquote{Spectroscopy by pulsed
  terahertz radiation,} {\protect\JournalTitle{Measurement Science and
  Technology}} \textbf{13}, 1727 (2002).

\bibitem{garnett1904xii}
J.~M. Garnett, \enquote{Xii. colours in metal glasses and in metallic films,}
  {\protect\JournalTitle{Philosophical Transactions of the Royal Society of
  London. Series A, Containing Papers of a Mathematical or Physical Character}}
  \textbf{203}, 385--420 (1904).

\bibitem{zhang2015effective}
X.~Zhang and Y.~Wu, \enquote{Effective medium theory for anisotropic
  metamaterials,} {\protect\JournalTitle{Scientific reports}} \textbf{5}, 7892
  (2015).

\bibitem{delrio2015fracture}
F.~W. DelRio, R.~F. Cook, and B.~L. Boyce, \enquote{Fracture strength of
  micro-and nano-scale silicon components,} {\protect\JournalTitle{Applied
  Physics Reviews}} \textbf{2} (2015).

\bibitem{gaither2013etching}
M.~S. Gaither, R.~S. Gates, R.~Kirkpatrick, \emph{et~al.}, \enquote{Etching
  process effects on surface structure, fracture strength, and reliability of
  single-crystal silicon theta-like specimens,} {\protect\JournalTitle{Journal
  of microelectromechanical systems}} \textbf{22}, 589--602 (2013).

\bibitem{kildishev2013planar}
A.~V. Kildishev, A.~Boltasseva, and V.~M. Shalaev, \enquote{Planar photonics
  with metasurfaces,} {\protect\JournalTitle{Science}} \textbf{339}, 1232009
  (2013).

\bibitem{jiang2018all}
X.~Jiang, H.~Chen, Z.~Li, \emph{et~al.}, \enquote{All-dielectric metalens for
  terahertz wave imaging,} {\protect\JournalTitle{Optics express}} \textbf{26},
  14132--14142 (2018).

\end{thebibliography}

\end{document}